\documentclass[12pt,twoside]{article}
\usepackage{amssymb}
\usepackage{amsmath}
\usepackage[dvips]{graphicx}
\begin{document}

\begin{center}
{\LARGE  Simulation of partial entanglement with one cbit and one M-box}

\vspace{0.6cm} {\bf Ali Ahanj}  and  {\bf Pramod S.
Joag}\footnote{\textsf{Electronic address:
pramod@physics.unipune.ernet.in}}

\vspace{0.2cm} {\it Department of Physics, University of Pune, Pune
- 411 007, India}

\vspace{0.4cm}

\end{center}

\begin{abstract}
We present a protocol to simulate the correlations implied by nonmaximally entangled two qubit states. We extend this protocol to simulate the non-local part of these correlations. These protocols use single cbit communication and a single use of Millionaire box (M-box). To the best of our knowledge, these resources are weaker than those used in previous  protocols using classical communication.      
\end{abstract}
\vspace{.2in}

\vspace{0.3cm}
\begin{flushright}
\end{flushright}
\textbf{I. INTRODUCTION}\\

One of the most intriguing features of quantum physics is the non-locality of correlations obtained by measuring entangled particles. These correlations are nonlocal because they are neither caused by an exchange of a signal, as any hypothetical signal should travel faster than light, nor are they due to any pre-determined agreement (shared randomness) as they break Bell's inequalities \cite{bell}.\\ A natural way to understand these correlations is to classically simulate them ( known as simulation of entanglement) using minimal resources. Obviously, this cannot be done using only local resources, that is, using shared randomness. The local resources must be supplemented by non-local ones. A simple non-local resource is communication of information via classical bits (cbits), which we can quantify and thus provides a `measure of non-locality'.\\In this scenario, Alice and Bob try and output $\alpha$ and $\beta$
respectively, through a classical protocol, with the same
probability distribution as if they shared the bipartite entangled
system and each measured his or her part of the system according to
a given random Von Neumann measurement. As we have mentioned above,
such a protocol must involve communication between Alice and Bob,
who generally share finite or infinite number of random variables. The amount of communication is quantified \cite{pironio03} either as
the average number of cbits $\overline {C}(P)$ over the directions
along which the spin components are measured (average or expected
communication) or the worst case communication, which is the maximum
amount of communication $C_{w}(P)$ exchanged between Alice and Bob
in any particular execution of the protocol. The third method is
asymptotic communication i.e., the limit
$lim_{n\rightarrow\infty}\overline{C}(P^{n})$ where $P^{n}$ is the
probability distribution obtained when $n$ runs of the protocol
carried out in parallel i.e., when the parties receive $n$ inputs
and produce $n$ outputs in one go.
In this paper we are concerned with the worst case scenario. A fundamental result for this scenario is that $k2^{n}$ ( $k$ a constant) cbits of classical communication is required to simulate the correlations implied by a  $n$ qubit maximally entangled state \cite{brassard}. This was followed by a remarkable result due to Toner and Bacon \cite{tonerbacon} who showed that a single cbit of communication is enough ( apart from shared random variables) to simulate the correlations of a two qubit singlet state. We have later shown \cite {A1}, \cite{A2} that for simulating entanglement of arbitrary spin $S$ singlet state, communication of $n=\log_{2}(s+1)$ cbits is enough, provided we confine only to spin measurements ( of type $\hat{a}\cdot\vec{S}$).\\

\textbf{II. SIMULATION USING NON-LOCAL BOXES}

Another fruitful approach to this problem is to use PR-box \cite{scarani1}. PR-box is a conceptual and mathematical tool developed to study non-locality, first proposed by Popescu and Rohrlich \cite{pr}. It was demonstrated in \cite{cerf}, that the correlations of the two qubit singlet can be simulated by supplementing hidden variables ( shared randomness) with a single use of the PR-box. Although mathematically based on the Toner and Bacon \cite{tonerbacon} result, this work is a major conceptual improvement, as the PR-box is a strictly weaker resource than a bit of communication, because it does not allow signaling.\\ However, it was soon discovered that a single use of PR-box is provably not sufficient to simulate some partially entangled two qubit states \cite{gisin}. Simulation of these states has been one of the open problems of the whole field. For example, \cite{wolf}, \cite{thesis} use a signaling resource strictly stronger than one cbit of communication. A reason for this apparently surprising result is that PR-boxes have random marginals, ($\langle \alpha\rangle=0=\langle \beta\rangle$, a fact which is consistent with two qubit singlet state) while the correlations arising from partially entangled quantum states have nontrivial marginals.\\Thus it appears that it is especially difficult to simulate at the same time nonlocal correlations and nontrivial marginals, like these corresponding to partially entangled quantum states. Recently, Brunner, Gisin, Popescu and Scarani \cite{bgps} (hereafter referred to as BGPS) have given a procedure to simulate entanglement in non-maximally entangled states using four PR-boxes and one M-box (see bellow). In order to overcome the above difficulty, they introduce the concept of correlated local flips, which is independent of whether we use nonlocal boxes or classical communication to simulate entanglement. The idea is that some nonlocal box, or a classical communication protocol, first simulates non-local correlations with trivial marginals and then use local flips to bias the marginals. In the following we outline the simulation scheme, based on local flips.\\The problem is to  simulate the quantum correlation implied by a general non-maximally entangled two-qubit state
\begin{equation}
\label{4}
|\psi(\gamma)\rangle=\cos (\gamma)|00\rangle+\sin (\gamma)|11\rangle ;~~~~~~~ (0\leq \gamma\leq\frac{\pi}{4}).
\end{equation}
 Alice and Bob perform measurements along directions 
$\hat{a}$ and $\hat{b}$ on their qubits ( that is, they measure operators $\vec{S}\cdot\hat{a}$ and $\vec{S}\cdot\hat{b}$). Let $\alpha, \beta$ $\left(\alpha , \beta \in\left\{-1,+1\right\}\right)$ denote their outputs respectively. The joint probability $P_{QM}(\alpha, \beta|\hat{a},\hat{b})$ is given by
\begin{equation}
\label{2}
 P_{QM}(\alpha, \beta|\hat{a},\hat{b})=\frac{1}{4}\big[1+\alpha c a_{z}+\beta c b_{z}+\alpha\beta C(\hat{a},\hat{b})\big],
 \end{equation}
 where $c\equiv\cos{2\gamma}$ and $s\equiv
\sin{2\gamma}$ and 
\begin{equation}
 C(\hat{a},\hat{b})=\langle \alpha \beta \rangle_{QM}=(a_{z}b_{z}+s(a_{x}b_{x}-a_{y}b_{y}))
 \end{equation}

In order to simulate this joint probability, we proceed as follows. We first set up a procedure ( involving non-local boxes or communication) to simulate the joint probability $P_0(\alpha, \beta|\hat{a},\hat{b})=\frac{1}{4}\big[1+\alpha\beta
C_0\big]$ where $C_0$ is the correlation $\langle\alpha \beta\rangle=\sum\alpha\beta P_0(\alpha, \beta|\hat{a},\hat{b}) $ {\it before} flipping.\\ Now, following BGPS, we invoke the local flip operation as follows. Alice (Bob) flips the output -1 with probability $f_{a}(f_{b})$ while the output +1 is left untouched. After the local flipping operation with probabilities $f_{a}$ and $f_{b}$ respectively by Alice and Bob, assuming $f_{b}\geq f_{a}$, the joint probability $P_{o}$ becomes
\begin{equation}
P_{f}(\alpha, \beta|\hat{a},\hat{b})=\frac{1}{4}\big[1+\alpha f_{a}+\beta f_{b}+ \alpha \beta(f_{a}+(1-f_{b})C_0)\big].
\end{equation}
In order that $P_{f}$ coincides with $P_{QM}$, we identify $f_{a}=ca_{z}$ and $f_{b}=cb_{z}$. Note that the condition $f_{b}\geq f_{a}$ now becomes $b_{z}\geq a_{z}$. This gives, $ C_{0}=\hat{a}\cdot\hat{B}$ where $\hat{B}=\left(sb_{x},sb_{y},b_{z}-c\right)/\left(1-cb_{z}\right)$. If $f_{a}\geq f_{b}$, $\left(a_{z}\geq b_{z}\right)$ , then
\begin{equation}
P_{f}(\alpha, \beta|\hat{a},\hat{b})=\frac{1}{4}\big[1+\alpha f_{a}+\beta f_{b}+ \alpha \beta(f_{b}+(1-f_{a})C_0)\big].
\end{equation}
where $ C_{0}=\hat{A}\cdot\hat{b}$ and $\hat{A}=\left(sa_{x},sa_{y},a_{z}-c\right)/\left(1-ca_{z}\right)$. It is easy to see that $\hat{A}$ and $\hat{B}$ are unit vectors. The joint probability $P_{0}$ to be simulated before local flipping operation depends on whether $\left(b_{z}\geq a_{z}\right)$ or $\left(a_{z}\geq b_{z}\right)$ ( via $C_{0}$). But Alice (Bob) cannot have any information on $b_{z}$ ($a_{z}$). In order to pave way through this situation BGPS invoke a nonlocal box called M-box which has two real inputs $ x,y  \in [0,1]$ and binary outputs $ m,n \in\left\{0,1\right\}$. The M-box is defined by $m\oplus n=[x\leq y]$ where $[x\leq y]$ is the truth value of the predicate $x\leq y $ for  given values of $x$ and $y$.\\

\textbf{III. THE PROTOCOL}

In this paper we give a protocol for simulating entanglement in an arbitrary non-maximally entangled two qubit quantum state, which uses single cbit of communication and a single use of M-box. The protocol runs as follows.\\ Alice and Bob share seven random unit vectors,  distributed independently and uniformly over unit sphere, $\hat{\lambda}_{1}$, $\hat{\lambda}_{2}$, $\hat{\mu}_{1}$
, $\hat{\mu}_{2}$, $\hat{\mu}_{3}$, $\hat{\mu}_{4}$, $\hat{\mu}_{5}$.  Also Alice and Bob share a common reference direction $\hat{z}$.\\ (i) Alice and Bob input $a_{z}$ and $b_{z}$ values respectively in M-box, which outputs $m$ and $n$ as above. We replace $m$ and $ n$ by $p$ and $ q$ respectively as $ p=2m-1$ and $ q=2n-1$ so that $ p,q \in\left\{-1,1\right\}$. Alice gets $p$ without knowing $q$ and Bob gets $q$ without knowing $p$.\\ (ii) Alice outputs
\begin{equation}
\alpha= \left(\frac{1+p}{2}\right)sgn\left(\hat{u}_{1}\cdot\hat{\lambda}_{1}\right)+ \left(\frac{1-p}{2}\right)sgn\left(\hat{u}_{2}\cdot\hat{\lambda}_{1}\right)
\end{equation}
where $\hat{u}_{1,2}=sgn\left(\hat{z}\cdot\hat{\mu}_{1,4}\right)\hat{a}+sgn \left(\hat{z}\cdot\hat{\mu}_{2,3}\right)\hat{A}+ \vec{a}_{1,2}$. Here $\vec{a}_{1,2}$ are vectors to make  $\hat{u}_{1,2}$ unit vectors.\\ (iii) Alice sends one cbit to Bob
\begin{equation}
c_{a}= \left(\frac{1+p}{2}\right)sgn\left(\hat{u}_{1}\cdot\hat{\lambda}_{1}\right)sgn\left(\hat{u}_{1}\cdot\hat{\lambda}_{2}\right)+ \left(\frac{1-p}{2}\right)sgn\left(\hat{u}_{2}\cdot\hat{\lambda}_{1}\right)sgn\left(\hat{u}_{2}\cdot\hat{\lambda}_{2}\right)
\end{equation}
(iv) Bob outputs
\begin{equation}
\beta= \left(\frac{1+q}{2}\right)sgn\left[\hat{v}_{1}\cdot\left(\hat{\lambda}_{1}+c_{a}\hat{\lambda}_{2}\right)\right]+ \left(\frac{1-q}{2}\right)sgn\left[\hat{v}_{2}\cdot\left(\hat{\lambda}_{1}+c_{a}\hat{\lambda}_{2}\right)\right]
\end{equation}
where $\hat{v}_{1,2}=sgn\left(\hat{z}\cdot\hat{\mu}_{3,2}\right)\hat{b}+sgn \left(\hat{z}\cdot\hat{\mu}_{1,4}\right)\hat{B}+ sgn\left(\hat{z}\cdot \hat{\mu}_{5}\right)\vec{b}_{1,2}$. Here $\vec{b}_{1,2}$ are vectors to make  $\hat{v}_{1,2}$ unit vectors.
Since $\int sgn\left(\hat{u}_{1,2}\cdot\hat{\lambda}_{1}\right)d\hat{\lambda}_{1}d\hat{\lambda}_{2}=0= \int sgn\left[\hat{v}_{1,2}\cdot\left(\hat{\lambda}_{1}+c_{a}\hat{\lambda}_{2}\right)\right]d\hat{\lambda}_{1}d\hat{\lambda}_{2}$, we see that at this stage of the protocol $ \left\langle \alpha \right\rangle_{0}= 0 =\langle\beta\rangle_{0}$. For $\left\langle \alpha \beta\right\rangle_{0}$ we have, after integrating over $\hat{\lambda}_{1,2}$
\begin{eqnarray}
\left\langle \alpha \beta\right\rangle_{\{\lambda\}}=\left(\frac{1+p+q+pq}{4}\right)\hat{u}_{1}\cdot\hat{v}_{1}+\left(\frac{1-p-q+pq}{4}\right)\hat{u}_{2}\cdot\hat{v}_{2}\nonumber \\+\left(\frac{1+p-q-pq}{4}\right)\hat{u}_{1}\cdot\hat{v}_{2}+\left(\frac{1-p+q-pq}{4}\right)\hat{u}_{2}\cdot\hat{v}_{1}
\end {eqnarray}
Next, integrating over $\{\hat{\mu}\}$ we get 
\begin{equation}
\label{1}
\left\langle \alpha \beta\right\rangle_{0}=\frac{1+pq}{2}\hat{a}\cdot\hat{B}+\frac{1-pq}{2}\hat{A}\cdot\hat{b}
\end{equation}
(v) Alice and Bob perform the local flip operation with probabilities $ f_{a}= c a_{z}$ and $f_{b}=c b_{z}$ respectively. Now, if $ f_{b} \geq f_{a} ( b_{z}\geq a_{z})$ then $p=q$ which implies from Eq.(\ref{1}) that $\left\langle \alpha \beta\right\rangle_{0}=\hat{a}\cdot\hat{B}=C_{0}$, so that $\left\langle \alpha \beta\right\rangle_{f}=ca_{z}+(1-cb_{z})\hat{a}\cdot\hat{B}=\left\langle \alpha \beta\right\rangle_{QM}$. If $ f_{a} \geq f_{b}( a_{z}\geq b_{z})$  then $p\neq q$ implying $\left\langle \alpha \beta\right\rangle_{0}=\hat{A}\cdot\hat{b}=C_{0}$, so that $\left\langle \alpha \beta\right\rangle_{f}=cb_{z}+(1-ca_{z})\hat{A}\cdot\hat{b}=\left\langle \alpha \beta\right\rangle_{QM}$.\\The local flipping operation ensures that $\left\langle \alpha \right\rangle_{f}=f_{a}=ca_{z}=\left\langle \alpha \right\rangle_{QM}$ and 
$\left\langle \beta \right\rangle_{f}=f_{b}=cb_{z}=\left\langle \beta \right\rangle_{QM}$. We see that the protocol simulates the joint probability $P_{QM}(\alpha, \beta|\hat{a},\hat{b}) $ as in Eq.(\ref{2}).\\

\textbf{IV. EPR2 DECOMPOSITION AND SIMULATION OF NON-LOCAL CORRELATION}

 Elitzur, Popescu and Rohrlich (EPR2)\cite{epr2} considered an experiment involving many photon pairs. The question they address is whether a subset of these pairs can be described with local correlations only, while the remaining ones are described non-locally. The global statistics, comprising the local and the non-local subsets of pairs, should reproduce the quantum statistics. Formally the EPR2 approach is to decompose the quantum correlations $P_Q$ as a convex sum of a local probability distribution $P_L$ and a non-local one $P_{NL}$ :
\begin{equation}
\label{3}
  P_Q=p_{L}(\rho)P_{L}+(1-p_{L}(\rho))P_{NL}.
\end{equation}
 The weight $p_{L}(\rho)$ is a measure of locality of the state $\rho.$  
 A decomposition of the form (\ref{3}) is particularly well-suited for the task of simulating quantum correlations, since only the non-local part $P_{NL}$ has to be simulated, the local part requiring only shared randomness. We use the improved version of the original EPR2 decomposition presented in \cite{scarani}. For the states $|\psi(\gamma)\rangle$ in Eq.(\ref{4}) we are interested in, the weight of the non-local part $p_{NL}(\gamma)=(1-p_{L}(\gamma))$ vanishes in the limit $\gamma\to 0$ of separable states. We present a simulation protocol for this $P_{NL}$  with one use of M-box and one cbit communication followed by correlated local flips. This second model, involving $P_L,P_{NL}$ decomposition, fulfills the desideratum that very weakly entangled states can be simulated by a vanishing amount of non-local resources. 
 The exact form of the decomposition described above, as given in \cite{scarani} is
 \begin{equation}
 \label{5}
 P_{NL}=\frac{1}{4}\Big[1+\alpha F(a_z)+\beta F(b_z)+\alpha\beta G(\hat{a},\hat{b})\Big]
 \end{equation}
  where $$F(x)=\frac{1}{s}\Big[cx-(1-s)f(x)\Big]$$ with $$f(x)=sgn(x) min\left(1,\frac{c}{1-s}|x|\right)$$ and 
  \begin{equation}
  \label{g}
  G(\hat{a},\hat{b})=a_x b_x-a_y b_y +\frac{1}{s}\left[a_z b_z- (1-s)f(a_z)f(b_z)\right]
  \end{equation}
   Scarani \cite{scarani} has shown that if both Alice's and Bob's measurement settings lie in a slice around the equator of the Bloch sphere, defined by 
  \begin{equation}
  \label{7}
  |a_z|,|b_z|\leq \frac{1-s}{c},
  \end{equation}
   $P_{NL}$ takes the simple form
   $\Big[F(a_z)=0= F(b_z)\Big]$ $$P_{NL}=\frac{1}{4}\Big[1+\alpha\beta(a_x b_x-a_y b_y-a_z b_z)\Big]=\frac{1}{4}\Big[1+\alpha\beta(\hat{a}\cdot{\hat{b}}^{\prime})\Big]$$  with ${\hat{b}}^{\prime}=(b_x,-b_y,-b_z).$ The local marginals are random and the correlations reduce to a simple scalar product, as in the singlet case. Further, if only one of the conditions (\ref{7}) is true, then the corresponding marginal is random and the local flip operation becomes redundant.\\ Note that, since $P_{QM}(-\alpha,\beta|-\hat{a},\hat{b})=P_{QM}(\alpha,\beta|\hat{a},\hat{b})$ it is sufficient to consider the case where $a_z,b_z\geq 0.$ Further, for each of the four possibilities corresponding to inequality (\ref{7}) it is straightforward to check that $\langle\alpha\beta\rangle_{F}=G(\hat{a},\hat{b})$ where $\langle\alpha\beta\rangle_{F}$ is the correlation after flipping operation. 
We now give a protocol using one use of M-box and one cbit of communication to simulate $P_{NL}$ of Eq.(\ref{5}) in all the four cases, namely,
\begin{eqnarray}
\label{4case}
  \left(a_z\leq \frac{1-s}{c}, b_z\leq \frac{1-s}{c}\right), \left(a_z\leq \frac{1-s}{c}, b_z > \frac{1-s}{c}\right)\nonumber\\ \left(a_z > \frac{1-s}{c}, b_z\leq \frac{1-s}{c}\right), \left(a_z > \frac{1-s}{c}, b_z > \frac{1-s}{c}\right).
 \end{eqnarray}
 Alice and Bob share 9 random unit vectors,  distributed independently and uniformly over unit sphere, $\hat{\lambda}_{1}$, $\hat{\lambda}_{2}$, $\hat{\mu}_{1}$
, $\hat{\mu}_{2}$, $\hat{\mu}_{3}$, $\hat{\mu}_{4}$, $\hat{\mu}_{5}$, $\hat{\mu}_{6}$, $\hat{\mu}_{7}$.\\
i) As in the first protocol Alice and Bob input $a_{z}$ and $b_{z}$ values respectively in M-box, whose outputs $p$ and $q$ are received by Alice and Bob respectively.\\
 (ii) Alice outputs
\begin{eqnarray}
\alpha &=& \left[a_z> \frac{1-s}{c}\right]\left[\left(\frac{1+p}{2}\right)sgn\left(\hat{u}_{1}\cdot\hat{\lambda}_{1}\right)+ \left(\frac{1-p}{2}\right)sgn\left(\hat{u}_{2}\cdot\hat{\lambda}_{1}\right)\right]\nonumber\\&+&
\left[a_z\leq  \frac{1-s}{c}\right]sgn\left(\hat{u}_{0}\cdot\hat{\lambda}_{1}\right)
\end{eqnarray}
where $$\hat{u}_{0}=\left[sgn\left(\hat{z}\cdot\hat{\mu}_{1}\right)+sgn\left(\hat{z}\cdot\hat{\mu}_{4}\right)+sgn\left(\hat{z}\cdot\hat{\mu}_{6}\right)\right]\hat{a}+ \vec{a}_{0}.$$ Here $\vec{a}_{0}$ is vector to make  $\hat{u}_{0}$ unit vector and  $u_{1,2}$ are the same as those in the first protocol except that  $\hat{A}$ is redefined as $\hat{A}=(sa_x,sa_y,c-a_z)/(1-ca_z).$\\ (iii) Alice sends one cbit to Bob
\begin{eqnarray}
c_{a} &=& \left[a_z> \frac{1-s}{c}\right]\left[\left(\frac{1+p}{2}\right)sgn\left(\hat{u}_{1}\cdot\hat{\lambda}_{1}\right)sgn\left(\hat{u}_{1}\cdot\hat{\lambda}_{2}\right)+ \left(\frac{1-p}{2}\right)sgn\left(\hat{u}_{2}\cdot\hat{\lambda}_{1}\right)sgn\left(\hat{u}_{2}\cdot\hat{\lambda}_{2}\right)\right]\nonumber\\
&+&\left[a_z\leq  \frac{1-s}{c}\right]sgn\left(\hat{u}_{0}\cdot\hat{\lambda}_{1}\right)
sgn\left(\hat{u}_{0}\cdot\hat{\lambda}_{2}\right)
\end{eqnarray}
(iv) Bob outputs
\begin{eqnarray}
\beta &=& \left[b_z>  \frac{1-s}{c}\right]\left[\left(\frac{1+q}{2}\right)sgn\left[\hat{v}_{1}\cdot\left(\hat{\lambda}_{1}+c_{a}\hat{\lambda}_{2}\right)\right]+ \left(\frac{1-q}{2}\right)sgn\left[\hat{v}_{2}\cdot\left(\hat{\lambda}_{1}+c_{a}\hat{\lambda}_{2}\right)\right]\right]\nonumber\\&+&\left[b_z\leq  \frac{1-s}{c}\right]sgn\left[\hat{v}_{0}\cdot\left(\hat{\lambda}_{1}+c_{a}\hat{\lambda}_{2}\right)\right]
\end{eqnarray}
where $$\hat{v}_{0}=\left[sgn\left(\hat{z}\cdot\hat{\mu}_{2}\right)+sgn\left(\hat{z}\cdot\hat{\mu}_{3}\right)+sgn\left(\hat{z}\cdot\hat{\mu}_{6}\right)\right]{\hat{b}}^{\prime}+sgn\left(\hat{z}\cdot\hat{\mu}_{7}\right) \vec{b}_{0}.$$ Here $\vec{b}_{0}$ is vector to make  $\hat{v}_{0}$ unit vector and  $\hat{v}_{1,2}$ are as in the first protocol except that $\hat{b}$ is replaced by ${\hat{b}}^{\prime}$ defined above. \\
After integrating over all $\lambda$ and $\mu$ we get 
\begin{eqnarray}
\label{18}
\langle \alpha \beta\rangle_{0} &=&\left[a_z>  \frac{1-s}{c}\right]\left[b_z>  \frac{1-s}{c}\right]\left\{\frac{1+pq}{2}\hat{a}\cdot\hat{B}+\frac{1-pq}{2}\hat{A}\cdot\hat{\acute{b}}\right\}\nonumber\\ &+& \left[a_z\leq  \frac{1-s}{c}\right]\left[b_z>  \frac{1-s}{c}\right]\hat{a}\cdot\hat{B}+\left[a_z>  \frac{1-s}{c}\right]\left[b_z\leq  \frac{1-s}{c}\right]\hat{A}\cdot{\hat{b}}^{\prime}\nonumber \\ &+&\left[a_z\leq  \frac{1-s}{c}\right]\left[b_z\leq  \frac{1-s}{c}\right]\hat{a}\cdot{\hat{b}}^{\prime}
\end{eqnarray}
(v) Alice (Bob) performs local flip operation with probability $F(a_z)$ ($F(b_z)$). Note that $F(a_z)=0$ ( $F(b_z)=0$)  whenever  $a_z\leq \frac{1-s}{c}$ ($b_z\leq \frac{1-s}{c}$) , so in this case,  Alice ( Bob) does not flip.  It is straightforward to show from Eq.(\ref{18}) that $ G(\hat{a},\hat{b})$ (Eq.(\ref{g})) equals $\langle \alpha \beta\rangle_{F}$ in all the four cases \ref{4case}. Since the local flipping with probabilities $F(a_z)=0$ and $F(b_z)=0$ produce required marginals, we see that our protocol simulates $P_{NL}$ ( Eq.(\ref{5})).  \\

\textbf{V. CONCLUSION}

In conclusion, we have presented protocols to simulate partially entangled two qubit states, using one cbit of communication and one use of M-box. Our work using EPR2 decomposition Eq.(\ref{3})  naturally confirms that the less the quantum state is entangled, less frequently one needs to use non-local resources as the simulation by BGPS shows. Further, the amount of non-local resources needed to simulate a partially entangled state is larger than that required to simulate maximally entangled states. The simulation of partial entanglement in BGPS uses all non-signaling resources [ PR and M-boxes] while our protocols use both the signaling and non-signaling resources. At any rate, our protocol is an improvement on previous protocols which used more than one cbit communication \cite{wolf}, as our protocol uses one cbit communication and one M-box which is a weaker resource. Since the existence of no signaling protocol and more than one cbit protocol does not imply the existence of a protocol with one cbit communication and one M-box, finding one such protocol is a new discovery in its own right and hopefully will find applications in future.\\

 {\textbf{ACKNOWLEDGMENT:}} We thank Guruprasad Kar
and Sibasish Ghosh for  useful discussions.

\end{document}